\documentclass[10pt]{llncs}
\usepackage[utf8]{inputenc}
\usepackage{amsmath}
\usepackage{amssymb}
\usepackage{xcolor}
\usepackage{a4wide}
\usepackage{tablefootnote}
\usepackage{epsfig}
\usepackage{listings}
\usepackage{graphicx}
\usepackage{enumitem}
\usepackage{float}
\usepackage{url}
\usepackage{cite}
\usepackage{mathtools}
\usepackage{booktabs}

\usepackage{geometry}
\title{Private LoRA Fine-tuning of Open-Source LLMs with Homomorphic Encryption}
\author{Jordan Frery, Roman Bredehoft,
        Jakub Klemsa, Arthur Meyre, Andrei Stoian}
 \institute{Zama} %, \\
\begin{document}

\maketitle

\begin{abstract}
Preserving data confidentiality during the fine-tuning of open-source Large Language Models (LLMs) is crucial for sensitive applications. This work introduces an interactive protocol adapting the Low-Rank Adaptation (LoRA) technique for private fine-tuning. Homomorphic Encryption (HE) protects the confidentiality of training data and gradients handled by remote worker nodes performing the bulk of computations involving the base model weights. The data owner orchestrates training, requiring minimal local computing power and memory, thus alleviating the need for expensive client-side GPUs. We demonstrate feasibility by fine-tuning a Llama-3.2-1B model, presenting convergence results using HE-compatible quantization and performance benchmarks for HE computations on GPU hardware. This approach enables applications such as confidential knowledge base question answering, private codebase fine-tuning for AI code assistants, AI agents for drafting emails based on a company's email archive, and adapting models to analyze sensitive legal or healthcare documents.
\end{abstract}

\section{Introduction}

Large Language Models (LLMs) exhibit transformative potential across a vast array of applications, from translation and summarization to classification and generation. While foundation models trained on web-scale datasets possess broad capabilities, unlocking their full potential for specific tasks or specialized domains like healthcare or law often requires fine-tuning. Fortunately, the prevalence of the transformer architecture allows many open-source LLMs to be efficiently adapted using techniques like Low-Rank Adaptation (LoRA) \cite{hu2022lora}.

However, adapting these powerful tools using sensitive, private data presents a significant challenge for organizations. Healthcare providers aiming to leverage confidential patient records, financial institutions using private transaction data, legal firms working with sensitive case documents, or companies customizing AI assistants on proprietary source code all face a critical dilemma. Exposing such raw data to third-party cloud services or even internal tools without stringent privacy guarantees is frequently untenable due to regulatory constraints (e.g., HIPAA, GDPR), competitive risks, or ethical considerations. Consequently, leveraging LLMs on valuable private datasets poses a major privacy problem.

While performing fine-tuning locally on client hardware avoids direct data exposure, this alternative presents its own significant challenges. State-of-the-art LLMs, even with efficient methods like LoRA, often demand substantial computational resources—particularly GPUs with high VRAM (e.g., $>$ 24GB)—which may exceed the capabilities or budget of many users or smaller organizations. Furthermore, establishing and managing the complex software environment for LLM fine-tuning requires considerable technical expertise. Therefore, outsourcing the computationally intensive parts of the fine-tuning process is highly desirable but demands a solution where data confidentiality can be rigorously maintained.

To address this critical need for \textbf{secure and accessible} private fine-tuning, we present a protocol where a client (data owner) interactively orchestrates LoRA fine-tuning of an open-source LLM, securely outsourcing the most demanding computations to a server (or a network of worker nodes). Our contribution lies in the \textbf{design and practical integration of a system enabling this private fine-tuning, centered around several key aspects}:
\begin{itemize}
    \item A client-server architecture tailored for LoRA, where the client manages private LoRA weights ($U, D$) and performs non-linear operations, while the server handles linear operations involving the public base model weights ($W$) under Homomorphic Encryption (HE).
    \item An efficient HE protocol for the core encrypted vector-clear matrix multiplication ($W \cdot [x]_\text{HE}$), utilizing packed Ring Learning With Errors (RLWE) ciphertexts for input/output, modulus switching for communication efficiency, and GPU acceleration via custom CUDA kernels.
    \item A demonstration of feasibility and convergence on a standard LLM (Llama-3.2-1B) using HE-compatible quantization.
\end{itemize}
In our setting, the base LLM is public, allowing the server to perform necessary computations, while the \textit{client exclusively holds the private LoRA weights representing the model's adaptation}. Efficiency is achieved by restricting server-side HE to only the linear operations involving $W$, leveraging optimized GPU kernels, and using strong ciphertext compression to minimize bandwidth. While the protocol naturally supports inference, our focus here is demonstrating the feasibility of \textit{private fine-tuning}.

\section{Prior Work}

\subsection{Low-Rank Adaptation}

LLM architectures are almost exclusively based on the multi-head attention (MHA) mechanism. For an input sequence $x$ (batch size $B$, context length $C$, model dimension $d$), a transformer layer computes updated representations $x'$ using MHA and a feed-forward network (FFN). Let $d_k$ be the dimension of keys and queries per attention head. The computation involves weight matrices ($W_Q, W_K, W_V, W_{proj}$) for attention and FFN weights (e.g., $W_{gate}, W_{up}, W_{down}$ in Llama-style models). Conceptually, for a single token (ignoring batch/context dimensions for simplicity):

\begin{equation}
    \begin{split}
        Q &= W_Qx, \quad K = W_Kx, \quad V = W_Vx \\
        \text{AttnOut} &= \operatorname{Attention}(Q, K, V) = \operatorname{softmax}\left(\frac{QK^T}{\sqrt{d_k}}\right)V \\
        h &= W_{proj}(\text{AttnOut}) \\
        \text{FFN}_{out} &= W_{down}(\operatorname{SiLU}(W_{gate}h) \odot W_{up}h) \\ % Llama-style FFN
        x' &= h + \text{FFN}_{out}
    \end{split}
    \label{eq:llm}
\end{equation}
Regular Low-Rank Adaptation (LoRA) fine-tuning \cite{hu2022lora} modifies a pre-trained weight matrix $W$ (size $d_{out} \times d_{in}$) by adding a low-rank update $\Delta W = UD$, where $U$ is $d_{out} \times r$ and $D$ is $r \times d_{in}$, with rank $r \ll \min(d_{in}, d_{out})$. The forward pass for a matrix multiplication $Wx$ is then modified to become $y = Wx + UDx (+ b \text{ if bias})$. For instance, applying LoRA to the query projection matrix $W_Q$ (where $d_{in}=d_{out}=d$) in Eq.~\eqref{eq:llm} changes the computation to $Q = W_Qx + U_Q D_Q x$. Only the LoRA matrices $U$ and $D$ are trained. LoRA is typically applied to attention weights ($W_Q, W_K, W_V, W_{proj}$) and FFN weights. The number of trainable parameters $r(d_{in} + d_{out})$ is much smaller than the original $d_{in}d_{out}$ parameters of $W$.

\subsection{Split Edge-Cloud LLM Fine-tuning}

Splitting LLM LoRA fine-tuning between an edge device and a cloud service has been explored in \cite{gao2024dlora, wang2023privatelora}. Both works outsource the computations involving the original model weight matrices $W$ to a cloud, while forward and backward passes on LoRA weights $U, D$ are kept local on the edge client. However, these approaches do not consider data confidentiality risks associated with sending intermediate activations to the cloud, and the latter work does not ensure model adaptation ($\Delta W$) confidentiality either.

\subsection{Non-interactive Encrypted Training}

In non-interactive training, only the encrypted dataset is sent to the server, and the encrypted model is retrieved. This has the advantage of keeping bandwidth requirements to a minimum. Logistic Regression training on encrypted data was described in several works \cite{10.1145/3643651.3659891, Kim2018, han2019logistic, bergamaschi2019homomorphic, Bonte2018}. Small multi-layer perceptrons (MLPs) were studied in \cite{10.1145/3643651.3659891, lou2020glyph, nandakumar2019towards}. Since HE operates over integers, most of these works quantize weights, gradients, activations, and the error function to between 6 and 8 bits. However, fully encrypted training requires a large amount of expensive encrypted multiplications and noise-management procedures like bootstrapping, making it too costly at present for large models like LLMs.

Outsourcing linear operations confidentially usually relies on HE (\cite{BatchCrypt}, \cite{yang2024packvflefficientpackingvertical}) or multi-party computation (MPC) \cite{mpc_dot_prod}. \cite{BatchCrypt} relies on the Paillier cryptosystem \cite{Paillier}. With Paillier, ciphertexts have sizes of 4096 or 6144 bits and can pack multiple plaintexts. Encryption is performed using modular exponentiation, though it can be optimized by splitting encryption into an online and an offline stage \cite{10.1007/978-3-031-30872-7_5}. When encrypting a large amount of data, pre-computing modular exponentiations in the offline stage may not be practical. Modular exponentiation makes encryption slow, placing a large computational burden on the client side and strongly reducing training throughput. RLWE-based approaches that pack multiple plaintexts in a single ciphertext are shown to be much more efficient than Paillier \cite{Gazelle}, both in terms of bandwidth needs and latency. \cite{Cheetah} introduces a protocol for private vector-matrix products with RLWE inputs and Learning With Errors (LWE) outputs, and \cite{hao2022iron} uses this protocol for LLM inference. These works do not address the training use case, which requires stronger data compression.

\section{Preliminaries}

\subsection{Homomorphic Encryption Scheme}\label{sec:HE}
Our approach utilizes an RLWE-based Homomorphic Encryption scheme~\cite{regev2005lwe,lyubashevsky2010rlwe}, similar to TFHE \cite{TFHE}. We briefly introduce the relevant concepts. Let $N$ denote the polynomial size (a power of 2, specified in Table~\ref{tbl:cryptoparams}).

\paragraph{RLWE Ciphertexts.} Let $\mathbb{Z}_p := \mathbb{Z}/p\mathbb{Z}$ (the ring of integers modulo $p$). An RLWE ciphertext encrypts a polynomial message $M \in R_p = \mathbb{Z}_p[X]/(X^N+1)$ under a secret key $S \in R_2$ (a polynomial with small coefficients). It is represented as a pair $(A, B) \in R_q \times R_q$, where $R_q = \mathbb{Z}_q[X]/(X^N+1)$, $q$ is the ciphertext modulus, $p$ is the plaintext modulus, $A$ is a uniformly random polynomial in $R_q$ called the \textbf{mask}, and $B = A \cdot S + E + \Delta M$. Here, $E \in R_q$ is a small noise polynomial (typically Gaussian), and $\Delta = q/p$ is a scaling factor. Decryption involves computing $B - A \cdot S \approx \Delta M$ and scaling down. The security relies on the hardness of the RLWE problem.

\paragraph{LWE Ciphertexts.} The Learning With Errors (LWE) problem is the basis for RLWE. An LWE ciphertext encrypts a single integer $m \in \mathbb{Z}_p$ under a secret key vector $\mathbf{s} \in \{0,1\}^n$. It is a pair $(\mathbf{a}, b) \in \mathbb{Z}_q^n \times \mathbb{Z}_q$, where $\mathbf{a}$ is a random vector (mask), and $b = \langle \mathbf{a}, \mathbf{s} \rangle + e + \Delta m$, with $e$ being small noise. Here, $n$ denotes the LWE dimension, which is related to the RLWE polynomial size $N$ in terms of security.

\paragraph{Mask Generation via PRNG.} To reduce communication overhead, the client and server can agree on a Pseudorandom Number Generator (PRNG). The client only sends a short seed $\mathbf{se}$, allowing both parties to deterministically generate the large mask polynomial $A$ locally. The client then only needs to transmit the seed and the computed body $B$, significantly compressing the ciphertext compared to sending both $A$ and $B$.

\subsection{Key HE Operations}
Our protocol relies on several standard HE primitives derived from \cite{TFHE}:
\begin{itemize}
    \item \textbf{SampleExtract:} Given an RLWE ciphertext $(A,B)$ encrypting polynomial $M$ and an index $h$, this operation extracts an LWE ciphertext $(\mathbf{a}',b')$ encrypting the $h$-th coefficient $M_h$ of $M$.
    \begin{equation}
    \textbf{SampleExtract}(\operatorname{RLWE}_S(M)_{A,B},h) \;\rightarrow\;
    \operatorname{LWE}_{S'}(M_h)_{\mathbf{a}',b'} := \begin{cases}
            a'_i = A_{h-i},      & 0 \le i \le h \\[2pt]
            a'_i = -A_{N+h-i},   & h <  i < N \\[2pt]
            b'   = B_h
    \end{cases}
    \end{equation}
    where $\mathbf{a}' = (a'_0, \dots, a'_{N-1})$ and $S'=(S_0,S_1,...S_{N-1})$, the coefficients of the RLWE secret key.

    \item \textbf{KeySwitching:} This operation changes the secret key under which a ciphertext is encrypted, typically from an LWE key $\mathbf{s}$ (dimension $n$) to an RLWE key $S'$ degree $N$. It is fundamental for packing multiple LWE ciphertexts into a single RLWE ciphertext. This requires a pre-computed Key Switching Key ($\textbf{KSK}$), which encrypts the bits of the original key $\mathbf{s}$ under the target key $S'$. In practice, $\textbf{KSK}$ is a specific set of RLWE ciphertexts. For an LWE ciphertext $(\mathbf{a}, b)$ encrypting $m$ under $\mathbf{s}$, the operation yields an RLWE ciphertext $(A', B')$ encrypting $m$ (or a constant term) under $S'$.
    \begin{equation}
    \mathbf{KeySwitching}(\operatorname{LWE}_{\mathbf{s}}(m)_{\mathbf{a}, b}, \textbf{KSK}) \rightarrow \operatorname{RLWE}_{S'}(m')_{(A',B')}
    \end{equation}
    is computed as:
    \begin{equation}
    (A', B') = (0, b) - \sum_{i=0}^{n-1} \textbf{Decomp}(a_i) \cdot \textbf{KSK}_{i}
    \end{equation}
    where $(0, b)$ represents a trivial RLWE encryption of $b$ in the constant term and $\textbf{KSK}_i$ is the ciphertext encrypting the $i$-th bit of the original key $\mathbf{s}$. Finally, the $\textbf{Decomp}$ algorithm is specific to the key switching procedure; for details, we refer to \cite{TFHE}.

    \item \textbf{ModulusSwitch:} This operation reduces the ciphertext modulus $q$ to a smaller modulus $q_{out}$. It serves to reduce the size of ciphertexts before they are transmitted back to the client, further lowering bandwidth.
    \item \textbf{Homomorphic Operations:} The scheme supports homomorphic addition and multiplication (by a cleartext value). We denote the homomorphic multiplication of an encrypted polynomial by a cleartext polynomial as $(\cdot)$.
    \item \textbf{Rotate:} Homomorphic rotation of the coefficients of the plaintext polynomial within an RLWE ciphertext. This rotation is negacyclic, meaning that coefficients wrapping around from the highest degree term acquire a sign change, corresponding to multiplication by $X^k$ in the quotient ring $\mathbb{Z}_p[X]/(X^N+1)$.
\end{itemize}
We refer the reader to \cite{TFHE} and related works for detailed algorithms and security analyses.

\section{Method}

\begin{figure*}[t]
    \centering
    \includegraphics[width=0.8\linewidth]{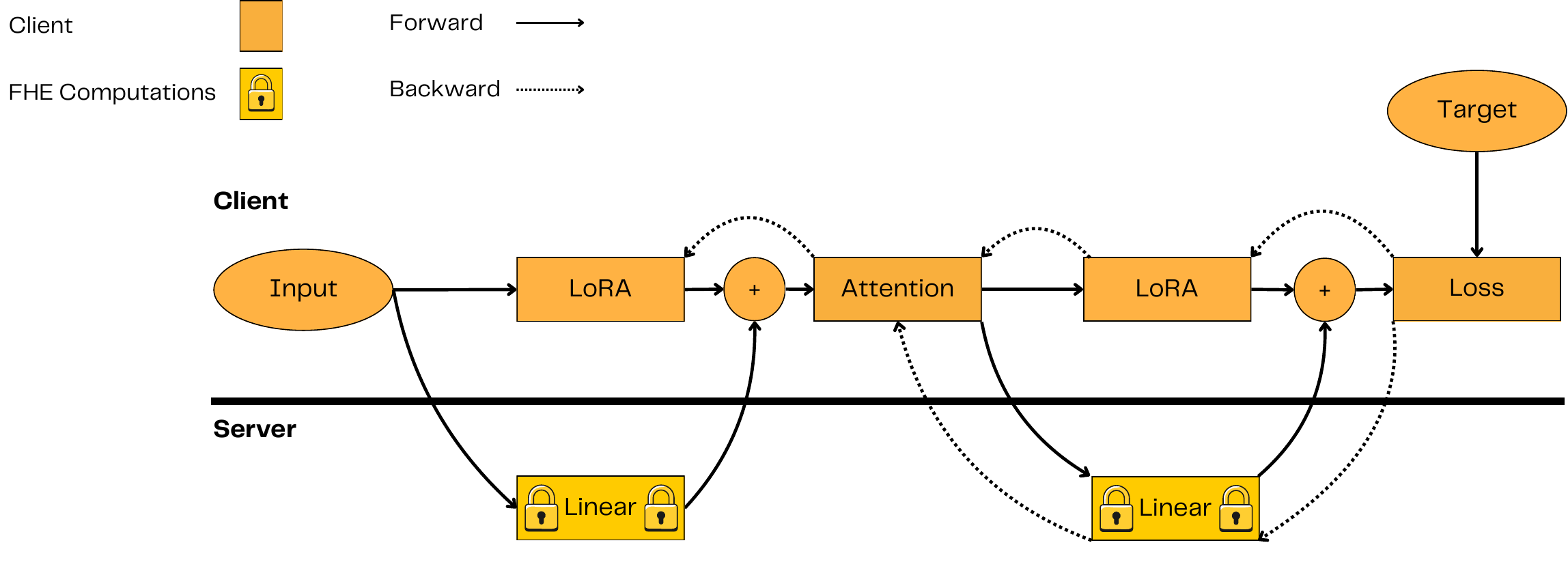}
    \caption{Private LoRA fine-tuning computation split: Client handles LoRA weights $U, D$ and non-linearities; Server handles base model weights $W$ under HE.}
    \label{fig:computation}
\end{figure*}

\subsection{LoRA Fine-tuning}
Figure~\ref{fig:computation} shows the responsibilities of the client and server in our LoRA fine-tuning protocol. The client performs local computations involving the private LoRA weights ($U, D$) and non-linear activation functions (softmax, SiLU). The server performs the computationally heavy linear operations involving the original known model weights ($W$) on homomorphically encrypted activations ($[x]_\text{HE}$). This split is expressed for a linear layer in Eq.~\eqref{eq:lora}.

\begin{equation}
    y = \underbrace{W \cdot [x]_\text{HE}}_\text{server-side} + \underbrace{UDx + b}_\text{client-side}
    \label{eq:lora}
\end{equation}
Here, $[x]_\text{HE}$ denotes the HE encryption of the activation $x$. The client receives the encrypted result $W \cdot [x]_\text{HE}$, decrypts it to obtain $Wx$, adds the locally computed $UDx + b$, applies the necessary activation function (if any), and re-encrypts the result for the next layer's server-side computation.

The matrices $D, U$ have shapes $(r, d), (d, r)$ respectively, where $d$ is the LLM hidden dimension and $r$ is the LoRA rank. The number of LoRA weights can thus be less than one percent of the number of weights in the original model. The weights $W$ are not updated with gradients, and the client only updates LoRA weights locally using any optimization strategy they see fit (e.g., Adam, AdaGrad). Attention modules that compute $h = \operatorname{softmax}(QK^T)V$ and the MLP activation are computed on the client side. The intermediate encrypted values needed for $Q,K,V$ (i.e., $W_Q[x]_\text{HE}$, $W_K[x]_\text{HE}$, $W_V[x]_\text{HE}$) and the inputs to the final client-side additions for $h$ and $x'$ are obtained from the server using the HE computation of Eq.~\eqref{eq:lora}.

\subsection{Quantization}
As LLMs typically operate on floating-point numbers and HE schemes work with integers, our approach requires quantization~\cite{jacob2018quantization}. This involves converting floating-point values $x_f$ to $n$-bit integers $x_q$.

\paragraph{Challenges with Standard LLM Quantization in HE.}
It is important to note that many popular quantization techniques developed for LLMs are not directly compatible with HE computation. Methods like block-wise quantization, where scaling factors are computed for small blocks of weights within a matrix, are primarily designed to reduce memory footprint and bandwidth during model loading \cite{dettmers2024qlora}. During computation (inference or training), these weights are often dequantized back to floating-point (e.g., bfloat16) on the fly to perform the matrix multiplications. This dequantization step is incompatible with HE, which fundamentally operates on encrypted integers. HE requires the entire computation, particularly the core matrix multiplications, to be performed using integer arithmetic on the quantized values. While some techniques like SmoothQuant \cite{xiao2023smoothquant} aim to make quantization more amenable to integer-only execution by migrating quantization difficulty from activations to weights, our current work focuses on simpler affine and symmetric quantization methods applied uniformly or with basic granularity, as detailed below.

\paragraph{Affine Quantization.}
We convert floating-point values $x_f$ to $n$-bit integers $x_q$ using a scaling factor $s_x$ and a zero-point $zp_x$: $x_f \approx s_x(x_q - zp_x)$. The parameters $s_x$ and $zp_x$ define the range and distribution of representable floating-point values. A common method to determine these is affine mapping based on the minimum ($x_{min}$) and maximum ($x_{max}$) values observed in the data:
\begin{align}
\begin{split}
    s_x &= \frac{x_{max} - x_{min}}{2^n - 1} \\
    zp_x &= \operatorname{round}\left( -\frac{x_{min}}{s_x} \right) \\
    x_q &= \operatorname{round}\left( \frac{x_f}{s_x} + zp_x \right)
\end{split}
\label{eq:uniformq}
\end{align}
The choice of $x_{min}$ and $x_{max}$ is critical and leads to different quantization strategies.

\paragraph{Static vs. Dynamic Quantization.}
The calculation of the range ($x_{min}, x_{max}$) can be done statically or dynamically.
\begin{itemize}
    \item \textbf{Static Quantization:} The range is determined once using a representative calibration dataset before inference or training. The same $s_x$ and $zp_x$ are then used for all subsequent inputs. This is simpler but may not capture the varying ranges encountered during execution.
    \item \textbf{Dynamic Quantization:} The range ($x_{min}, x_{max}$) is computed \textit{on-the-fly} for each input tensor based on its actual values. This adapts better to varying data distributions but incurs runtime overhead for range calculation.
\end{itemize}
In our HE context, dynamic quantization parameters for activations sent to the server must be computed by the client and also sent. Static parameters for weights are fixed beforehand.

\paragraph{Quantization Granularity.}
The quantization parameters ($s_x, zp_x$) can be applied at different levels of granularity within a tensor:
\begin{itemize}
\item \textbf{Per-Tensor:} A single set of parameters ($s_x, zp_x$) is used for the entire weight or activation tensor. This is the simplest approach.
\item \textbf{Per-Channel:} For weight tensors (e.g., shape \texttt{[out\_channels, in\_channels]}), separate parameters are computed for each output (or input) channel. This better accommodates varying weight scales across different neurons.
\item \textbf{Per-Token:} For activation tensors (e.g., shape \texttt{[batch, sequence\_len, features]}), separate parameters can be computed for each token along the sequence length dimension. This captures fine-grained variations in activation magnitudes within a sequence.
\end{itemize}
Finer granularity generally improves accuracy but increases the number of scale/zero-point parameters to manage.

\paragraph{Symmetric Quantization for HE.}
Substituting the affine quantizers
$x_f \approx s_x\,(x_q - zp_x)$ and
$w_f \approx s_w\,(w_q - zp_w)$
into the inner product $\sum x_f w_f$ (cf. Eq.~\eqref{eq:uniformq}) gives the integer expression
\[
\sum_{i=1}^{K} x_q^{(i)} w_q^{(i)}
\;-\;
zp_w \sum_{i=1}^{K} x_q^{(i)}
\;-\;
zp_x \sum_{i=1}^{K} w_q^{(i)}
\;+\;
K\,zp_x zp_w ,
\]
where $K$ is the number of accumulated terms (e.g., the input dimension $d_{in}$). If the zero-points are non-zero, all four sums must be evaluated homomorphically—which is costly under HE. We therefore adopt \textbf{symmetric quantization}: choose the ranges such that $x_{\min}\approx -x_{\max}$ and $w_{\min}\approx -w_{\max}$, which forces \(zp_x = zp_w = 0\) (for signed integers). With the offset terms eliminated, the server's task reduces to computing only the encrypted dot product $\sum_i x_q^{(i)} w_q^{(i)}$, after which the client re-applies the cleartext scale factor $s_x s_w$. Both weights and activations are quantized symmetrically in our HE protocol.

Our experiments (Sec.~\ref{sec:eval_quant_convergence}) explore the impact of these different strategies (static vs. dynamic, granularity) on model convergence.

\subsection{Encrypted Vector -- Clear Matrix Multiplication}
The main objective for the encrypted vector -- clear matrix multiplication ($W \cdot [x]_\text{HE}$) design was to keep ciphertext sizes low while maintaining HE performance.

\subsubsection{Computation Method}
We split the input activation vector $x$ (of size $d_{in}$) into $L = \lceil d_{in}/N \rceil$ blocks, $\hat{x}_0,\hat{x}_1,\dots,\hat{x}_{L-1}$, padding the last block with zeros if necessary. The client encrypts each block individually using the RLWE scheme, obtaining $\operatorname{RLWE}(\hat{x}_0)$, \dots, $\operatorname{RLWE}(\hat{x}_{L-1})$. The server holds the cleartext weight matrix $W$, whose columns $w_j$ are conceptually split into corresponding blocks $\hat{w}_{0j}, \dots, \hat{w}_{L-1,j}$, where each $\hat{w}_{ij}$ aligns with the block $\hat{x}_i$.

To obtain the dot-product between the encrypted input $[x]_\text{HE}$ and a cleartext matrix column $w_j$, the server performs the following computation using homomorphic operations:
\begin{equation}
\operatorname{LWE}(x\cdot w_j) = \sum_{i=0}^{L-1} \textbf{SampleExtract}(\operatorname{RLWE}(\hat{x}_i)\cdot \hat{w}_{ij}, N-1)
\label{eq:LWEdot}
\end{equation}

Here, $\operatorname{RLWE}(\hat{x}_i)\cdot \hat{w}_{ij}$ represents the homomorphic multiplication of the encrypted block $\operatorname{RLWE}(\hat{x}_i)$ by the cleartext polynomial block $\hat{w}_{ij}$. Weights are encoded in reverse order (i.e., $\hat{w}_{ij}[k] = w_{j}[iN+N-1-k]$) for efficient dot product computation via the highest coefficient of the polynomial multiplication. \textbf{SampleExtract} then isolates the LWE encryption of the $i$-th partial sum of the dot product.

The final \textbf{ModulusSwitch} to a smaller $q_{out}$ compresses the output ciphertext that is to be sent back to the client.

Finally, the resulting LWE samples $\operatorname{LWE}(x\cdot w_j)$ for all columns $j$ (from $j=0$ to $d_{out}-1$) are efficiently packed back into a single output RLWE ciphertext using \textbf{KeySwitching} and homomorphic rotations (\textbf{Rotate}), conceptually represented as:
\begin{equation}
    \operatorname{RLWE}(Wx) = \textbf{ModulusSwitch}_{q_{out}}\left(\sum_{j=0}^{d_{out}-1}\textbf{Rotate}(\textbf{KeySwitching}(\operatorname{LWE}(x\cdot w_j)), j)\right)
    \label{eq:pack_rotate}
\end{equation}

We use coefficient packing into RLWE ciphertexts (defined in Eq.~\eqref{eq:pack_rotate}). Our approach is similar to \cite{Cheetah} but adds steps for enhanced communication efficiency: (1) careful packing of intermediate dot-products, and (2) final \textbf{ModulusSwitch} of the output RLWE ciphertext.

\subsection{Cryptosystem Parameters}

We require that the encoding precision $\beta$ prevents overflows during computation for any $x$ or $W$, provided the quantization protocol is followed. Furthermore, we allow noise to impact $\beta - \gamma$ least significant bits (LSBs) of the dot-product result. Since, for layer $k+1$, the client sends a re-quantized result derived from the decrypted $k$-th linear layer output, the full precision of the $k$-th layer result is not strictly necessary.

Following the constraints above, the cryptosystem parameters are chosen to provide 128-bit security according to the lattice estimator, and their values are given in Table~\ref{tbl:cryptoparams}.

\begin{table}[H]
\centering
\caption{Cryptosystem parameters.}
\label{tbl:cryptoparams}
\begin{tabular}{@{}lll@{}}
\toprule
Parameter        & Description                       & Value                 \\ \midrule
$\beta$          & Bits reserved for computation     & 27                    \\
$\gamma$          & MSBs unaffected by noise growth    & 12                    \\
N                & Polynomial size                 & 2048                  \\
$q_{in}$         & Input modulus (bits)              & 39                    \\
$q_{out}$        & Output modulus (bits)             & 26                    \\
$\sigma_{input}$ & Input noise distribution std. dev. & 2.845e-15  \\
$\sigma_{ksk}$   & Keyswitching key noise std. dev.   & 2.845e-15 \\ \bottomrule
\end{tabular}
\end{table}

Considering these parameters, we can compute the communication expansion factor. We define this as the total size of the input/output ciphertexts divided by the size of the corresponding plaintext values (assuming 8-bit quantization). Let $N = 2048$.

The expansion factor is calculated as follows:
\begin{enumerate}
    \item Input $x$: The client sends a seed ($\mathbf{se}$, 64 bits = 8 bytes) and the RLWE body $B$ which has $N$ coefficients with modulus $q_{in}$ (39 bits). Size of $B \approx N \times q_{in} / 8 = 2048 \times 39 / 8 \approx 9984$ bytes. Total size $\approx 8 + 9984 = 9992$ bytes. This encrypts $N=2048$ plaintext values. Assuming 8-bit plaintext values (1 byte each), the input plaintext size is 2048 bytes. Expansion factor (Input): $9992 / 2048 \approx 4.88$.
    \item Output $Wx$: The client receives the full RLWE ciphertext $(A', B')$. Both $A'$ and $B'$ have $N=2048$ coefficients with modulus $q_{out}$ (26 bits). The size of each component is $N \times q_{out} / 8 = 2048 \times 26 / 8 \approx 6656$ bytes. The total transmitted size is therefore $2 \times 6656 = 13312$ bytes. This ciphertext represents $N=2048$ output values resulting from the homomorphic computation. Since the scheme guarantees $\gamma=12$ correct MSBs per value, the effective plaintext information size is $N \times \gamma / 8 = 2048 \times 12 / 8 = 3072$ bytes. Expansion factor (Output): $13312 / 3072 \approx 4.33$.
\end{enumerate}

\subsection{GPU Implementation}
\label{sec:gpu}
Matrix multiplication is a core computation performed on GPUs, used extensively in machine learning, particularly in LLMs. While common \textbf{MatMul} implementations work on floating-point numbers, for this work, we implemented one for 64-bit and 32-bit integers based on \cite{GEMM}. On an NVIDIA RTX 4060 Laptop GPU, our kernel achieves approximately $800 \times 10^9$ integer operations per second, roughly 4x slower than the floating-point reference implementation in the \textbf{cuBLAS} library.

We implement the various parts of the encrypted vector-matrix computation using GPU kernels, leveraging the integer \textbf{MatMul}. We consider a batch of input vectors. The computation in Eq.~\eqref{eq:LWEdot} involves homomorphic polynomial multiplication ($\operatorname{RLWE}(\hat{x}_i)\cdot \hat{w}_{ij}$), which is computed coefficient-wise, followed by \textbf{SampleExtract} and summation to produce LWE ciphertexts. The subsequent packing step via \textbf{KeySwitching} (part of Eq.~\eqref{eq:pack_rotate}) relies heavily on matrix multiplications and is well-suited for GPU acceleration.

For a given input vector $x$, the computation in Eq.~\eqref{eq:LWEdot} yields one LWE ciphertext $\operatorname{LWE}(x \cdot w_j) = (\mathbf{a}_j, b_j)$ for each column $j$ of the weight matrix $W$ (from $j=0$ to $d_{out}-1$), where $\mathbf{a}_j \in \mathbb{Z}_q^N$ is the LWE mask vector and $b_j \in \mathbb{Z}_q$ is the LWE body. To perform KeySwitching efficiently for all columns in parallel, we aggregate these LWE components. Let $A_{LWE}$ be the matrix of size $d_{out} \times N$ whose rows are the mask vectors $\mathbf{a}_j$, and let $b_{LWE}$ be the column vector of size $d_{out}$ containing the bodies $b_j$. The KeySwitching operation, transforming these $d_{out}$ LWE ciphertexts (under key $\mathbf{s}$) into $d_{out}$ RLWE ciphertexts (under key $S'$), can then be expressed using matrix products suitable for the \textbf{MatMul} kernel:
\begin{equation}
    \begin{bmatrix}
        \operatorname{RLWE}(x \cdot w_0) \\
        \vdots \\
        \operatorname{RLWE}(x \cdot w_{d_{out}-1})
    \end{bmatrix}
        = \begin{cases}
        A_{RLWE} = 0 - \operatorname{MatMul}\left(\textbf{Decomp}(A_{LWE}) , \textbf{KSK}_A\right)
        ,\\
        B_{RLWE} = b_{LWE} -
        \operatorname{MatMul}\left(\textbf{Decomp}(A_{LWE}) , \textbf{KSK}_B\right)
    \end{cases}
    \label{eq:gpu_keyswitch}
\end{equation}
Here, $\textbf{Decomp}(A_{LWE})$ represents the matrix resulting from applying the decomposition algorithm (see \textbf{KeySwitching} in Section~\ref{sec:HE}) to the LWE masks. $\textbf{KSK}_A$ and $\textbf{KSK}_B$ are components derived from the Key Switching Key ($\textbf{KSK}$), pre-computed based on the source LWE key $\mathbf{s}$ and target RLWE key $S'$. This matrix formulation allows the computationally intensive part of KeySwitching to be performed efficiently using the GPU's \textbf{MatMul} kernel. The resulting $A_{RLWE}$ and $B_{RLWE}$ contain the $d_{out}$ RLWE ciphertexts, where the $j$-th ciphertext encrypts $x \cdot w_j$.
In the subsequent step, each resulting $\operatorname{RLWE}(x \cdot w_j)$ ciphertext is homomorphically rotated by degree $j$, and the rotated ciphertexts are summed together, as per Eq.~\eqref{eq:pack_rotate}. This final result is modulus switched, packed into a bit-vector, and returned to the client.
\subsection{Client-Side Computation Sizing}
\label{sec:client_sizing}

In our protocol, the client performs computations involving the LoRA adapters ($U, D$) and non-linearities. We estimate the client's computational load (FLOPs) per layer for a forward and backward pass over a context of $C$ tokens (batch size $B=1$). Let $d$ be the hidden dimension, $m$ the intermediate FFN dimension, $r$ the LoRA rank, $d_k$ the key/query dimension per head, $d_v$ the value dimension per head, and $n_{layers}$ the number of layers.

\begin{enumerate}
    \item \textbf{Attention Mechanism:} The client computes $QK^T$ and the subsequent product with $V$ after receiving decrypted $W_Q[x]_\text{HE}$, $W_K[x]_\text{HE}$, $W_V[x]_\text{HE}$ from the server and adding the local LoRA contributions. The $QK^T$ operation involves matrices of size $C \times d_k$ and $d_k \times C$ (per head), resulting in $C \times C$ attention scores. Summing over $n_{heads}$ (where $d = n_{heads}d_k$), this requires approximately $d C^2$ FLOPs. The product of the $C \times C$ attention scores with $V$ (size $C \times d_v$, where $d = n_{heads}d_v$) requires another $d C^2$ FLOPs. Total non-LoRA client MHA FLOPs per layer $\approx 2dC^2$.
    \item \textbf{LoRA Adapter Computation ($UDx$):} Computing $UDx$ for a layer with input dimension $d_{in}$ and output dimension $d_{out}$ requires calculating $Dx$ ($r \times d_{in}$ multiplications, $r \times (d_{in}-1)$ additions $\approx 2 r d_{in}$ FLOPs) and then $U(Dx)$ ($d_{out} \times r$ multiplications, $d_{out} \times (r-1)$ additions $\approx 2 r d_{out}$ FLOPs), totaling approximately $2r(d_{in} + d_{out})$ FLOPs per adapter application.
    \item \textbf{LoRA FLOPs per Layer:} We assume LoRA is applied to $W_Q, W_K, W_V, W_{proj}$ (all $d_{in}=d, d_{out}=d$) and the FFN layers. For Llama-style FFNs, this includes $W_{gate}, W_{up}$ ($d_{in}=d, d_{out}=m$) and $W_{down}$ ($d_{in}=m, d_{out}=d$).
        \begin{itemize}
            \item Attention LoRA (Q, K, V, Proj): $4 \times 2r(d + d) = 16dr$ FLOPs.
            \item FFN LoRA (Gate, Up, Down): $2 \times 2r(d + m) + 2r(m + d) = 6rd + 6rm$ FLOPs.
            \item Total LoRA FLOPs per layer (forward pass): $16dr + 6dr + 6mr = 22dr + 6mr$ FLOPs.
        \end{itemize}
\end{enumerate}

Combining these for a single forward pass per layer gives approximately $2dC^2 + 22dr + 6mr$ FLOPs. Assuming the backward pass requires roughly twice the FLOPs of the forward pass (a common rule of thumb), the total client-side computation for one context of $C$ tokens across all $n_{layers}$ is on the order of:

\begin{equation}
    \textbf{Client}_{FLOPs} \approx 2 \times n_{layers} (2dC^2 + 22dr + 6mr) \quad \textbf{FLOPs}
    \label{eq:client_comp}
\end{equation}

\section{Security Model}

This section outlines the security properties and assumptions of our private LoRA fine-tuning protocol.

\subsection{Threat Model}
We consider the server executing the HE computations as \textbf{\textit{honest-but-curious}}. This means the server correctly follows the protocol but may attempt to infer information about the client's private data from the encrypted messages exchanged. We do not consider malicious servers who actively deviate from the protocol (e.g., by tampering with computations); protection against such adversaries is beyond the scope of this work.

\subsection{Protected Assets}
The protocol aims to protect the confidentiality of the following client-owned assets from the server:
\begin{itemize}
    \item The raw fine-tuning data (e.g., text inputs, labels).
    \item Intermediate activations and gradients derived from the private data that are processed homomorphically by the server.
    \item The final trained LoRA weights $\Delta W = UD$, representing the client's private model adaptation.
\end{itemize}

\subsection{Assumptions}
The security of the protocol relies on the following assumptions:
\begin{itemize}
    \item The base LLM weights $W$ are publicly known or available to both client and server.
    \item The underlying RLWE-based HE scheme provides semantic security (IND-CPA), ensuring ciphertexts do not reveal information about the plaintexts.
    \item The client machine is secure and trusted.
    \item Cryptographic parameters (Table~\ref{tbl:cryptoparams}) provide a standard level of security (e.g., 128-bit) against known attacks.
\end{itemize}

\subsection{Security Guarantees}
Under the honest-but-curious threat model and the stated assumptions, the protocol ensures that the server learns no information about the client's private data or the resulting LoRA weights $\Delta W$. The semantic security of HE protects all data processed homomorphically on the server. Furthermore, the LoRA weights $U$ and $D$ are managed exclusively on the client side and never shared, encrypted or otherwise, with the server. Potential information leakage is limited to data-independent side channels like computation counts or timing patterns, which are not explicitly addressed here.

\section{Evaluation}

We evaluate our private LoRA fine-tuning approach using experiments conducted with the Llama-3.2-1B \cite{llama32024} open-source model (1B parameters, $n_{layers}=16$, hidden size $d=2048$, FFN intermediate size $m=8192$) on various tasks. All experiments were performed simulating a client interacting with a single server. Our evaluation focuses on demonstrating:
\begin{enumerate}[label=(\roman*)]
    \item The feasibility and convergence of LoRA fine-tuning using 8-bit quantization compared to floating-point.
    \item The correctness of the HE execution path by comparing loss trajectories between the quantized cleartext execution and the HE execution setting.
    \item The performance (timing) of the actual HE execution on a representative task using the Llama-3.2-1B model in our single-server setup, including the resulting client compute rate.
    \item The qualitative impact of fine-tuning on model outputs.
\end{enumerate}
All experiments utilized components from the Concrete ML \footnote{Implementation available in the Concrete ML library: \url{https://github.com/zama-ai/concrete-ml}}
 library, implementing the client-server computation split described in Section 4. Unless otherwise specified, experiments used 8-bit symmetric quantization, with LoRA rank $r=8$ and $\alpha=32$.

\subsection{Correctness and Convergence in Quantized Cleartext}
\label{sec:eval_quant_convergence}

We carried out an extensive ablation study in \emph{cleartext} to establish which quantization policies allow LoRA fine-tuning to converge reliably at low bit-widths. Table~\ref{tab:quant_settings_overview} gives an overview of the concrete settings we evaluated, which cover combinations of:

\begin{itemize}[nosep]
  \item \textbf{Range selection}: \emph{static} (\texttt{S}) vs. \emph{dynamic} (\texttt{D}) (computed per input tensor),
  \item \textbf{Granularity}: per-tensor (\texttt{T}), per-token for activations (\texttt{Tok}), and per-channel for weights (\texttt{C}),
  \item \textbf{Bit-width}: 8, 16, and the FP32 reference.
\end{itemize}

A shorthand notation of the form \textit{Activation-Weight} (e.g., \texttt{DTok-SC}) denotes the pair of strategies applied to activations and weights, respectively. For instance, \texttt{DTok} means \emph{D}ynamic range on a per-\emph{Tok}en basis, whereas \texttt{SC} stands for \emph{S}tatic range on a per-\emph{C}hannel basis.

\begin{figure}[H]
  \centering
  \includegraphics[width=0.85\linewidth]{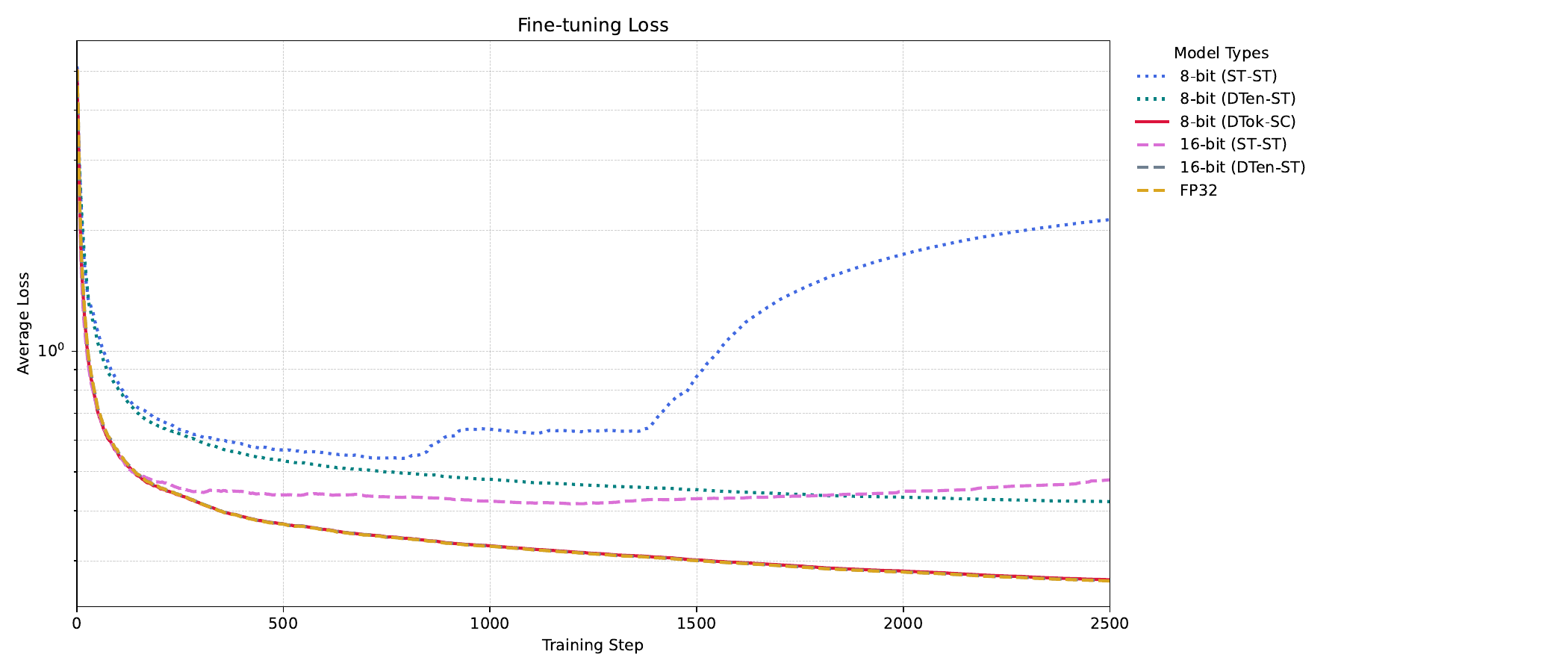}
  \caption{Training-loss trajectories for quantization settings in Table~\ref{tab:quant_settings_overview} on Orca Math.}
  \label{fig:quant_loss_strategies_full}
\end{figure}

\paragraph{Experimental protocol.} All runs fine-tuned Llama-3.2-1B for 2500 training steps on the \textsc{Orca-Math} dataset using rank $r=8$ LoRA adapters ($\alpha=32$). Each model used the same initial optimizer, weights, and data state to ensure comparable learning dynamics. Convergence was monitored through the running average of the training loss, and the quantitative results are summarized in Figure~\ref{fig:quant_loss_strategies_full}. We observe the following:

\begin{enumerate}[label=(\alph*),nosep]
  \item \textbf{Static per-tensor (\texttt{ST-ST}) struggles significantly at low bit-widths.} At 8-bit, the loss initially converged for approximately 700 steps before exhibiting instability, marked by sudden increases—a sign of potential gradient explosion due to the limited dynamic range captured by a single static scale. At 16-bit, \texttt{ST-ST} performed better, converging to a lower loss initially, but the loss eventually flattened and began increasing around step 2000, indicating that even at 16-bit, a static per-tensor approach can be suboptimal for capturing the full dynamics of training.
  \item \textbf{Dynamic range is crucial, especially for activations.} Introducing dynamic range for activations (\texttt{DT-ST}) at 8-bit prevented the gradient explosion seen with \texttt{ST-ST}. The loss converged steadily after the initial 300 steps, although it still settled at a higher value (0.42) than the FP32 reference. This shows that while dynamic range helps stabilize training, per-tensor granularity at 8-bit remains insufficient to fully match FP32 convergence. Gradients, in particular, benefit from dynamic range estimation as their magnitude can change dramatically during training.
  \item \textbf{16-bit precision is more forgiving regarding granularity.} The \texttt{DT-ST} setting at 16-bit achieved convergence nearly identical to the FP32 baseline (final loss 0.29). This highlights that with sufficient bit-width (16-bit), even simpler granularity strategies (dynamic per-tensor for activations, static per-tensor for weights) can yield excellent results, validating the underlying quantization implementation.
  \item \textbf{Fine granularity unlocks 8-bit performance near FP32.} To match FP32 convergence at 8-bit, finer granularity is necessary. Introducing per-channel static weights (\texttt{DT-SC}) or per-token dynamic activations (\texttt{DTok-ST}) significantly improved 8-bit performance (final loss 0.29 and 0.30, respectively). Combining both—dynamic per-token activations and static per-channel weights (\texttt{DTok-SC})—yielded an 8-bit loss trajectory virtually indistinguishable from the FP32 reference (final loss 0.27).
  \item \textbf{16-bit serves as a virtually lossless upper bound.} All granularity variants tested at 16-bit (including the most granular \texttt{DTok-SC}) converged within numerical noise of FP32, confirming that 16-bit quantization is a robust and easy choice when near-lossless accuracy is required.
\end{enumerate}

 \paragraph{Final Perplexity Confirmation.} We further confirmed these findings by evaluating the perplexity on the Orca-Math validation set after 2500 training steps. Table~\ref{tab:quant_perplexity} shows the results. The 8-bit \texttt{DTok-SC} configuration achieved a perplexity (1.2391) nearly identical to the FP32 baseline (1.2381). Conversely, the \texttt{ST-ST} configurations yielded significantly higher perplexity values, agreeing with their poor performance observed in the loss curves.

 \begin{table}[H]
   \centering
   \caption{Final perplexity on Orca Math validation set after 2500 training steps for different quantization settings. Lower is better. Best performing configurations are highlighted.}
   \label{tab:quant_perplexity}
   \begin{tabular}{@{}lc@{}}
     \toprule
     \textbf{Configuration} & \textbf{Final Perplexity} \\ \midrule
     FP32               & \textbf{1.2381}       \\
     8-bit (DTok-SC)    & \textbf{1.2391}       \\
     16-bit (DT-ST)     & 1.2506           \\
     8-bit (DT-ST)      & 1.4898           \\
     8-bit (ST-ST)      & 33.4061          \\
     16-bit (ST-ST)     & 44.6870          \\ \bottomrule
   \end{tabular}
 \end{table}

 This study reveals that \emph{dynamic range} calculation, particularly for activations (and implicitly, gradients), is essential for stable low-bit training. Furthermore, achieving convergence close to FP32 at very low bit-widths like 8-bit necessitates \emph{fine-grained} quantization strategies. The \texttt{DTok-SC} recipe consistently delivered near-float accuracy at 8 bits, confirmed by both loss and perplexity metrics, and was therefore adopted as the default for our HE experiments (Sections~\ref{sec:fhe_performance}–\ref{sec:qualitative}).

\begin{table}[H]
  \centering
  \caption{Quantization schemes explored in the cleartext study on Orca Math. The two leftmost columns specify the range selection and granularity for activations and weights; the remaining columns report the final training loss after one epoch at 8- and 16-bit. Lower is better.}
  \label{tab:quant_settings_overview}
  \begin{tabular}{@{}llcc@{}}
    \toprule
    \textbf{Activations} & \textbf{Weights} & \textbf{8-bit} & \textbf{16-bit} \\ \midrule
    ST  & ST  & 2.1 & 0.29 \\
    DT  & ST  & 0.42 & 0.29 \\
    DTok & SC  & \textbf{0.27} & \textbf{0.27} \\
    \bottomrule
  \end{tabular}
\end{table}

\subsection{HE Execution Fidelity}
\label{sec:fhe_execution_fidelity}
To verify that our homomorphic back-end faithfully reproduces cleartext training dynamics while preserving numerical precision, we conducted two complementary evaluations: (1) \emph{training-loss fidelity} on Llama-3.2-1B, and (2) \emph{bit-level error analysis} of decrypted dot-product computations.

\begin{figure}[H]
  \centering
  \includegraphics[width=0.75\linewidth]{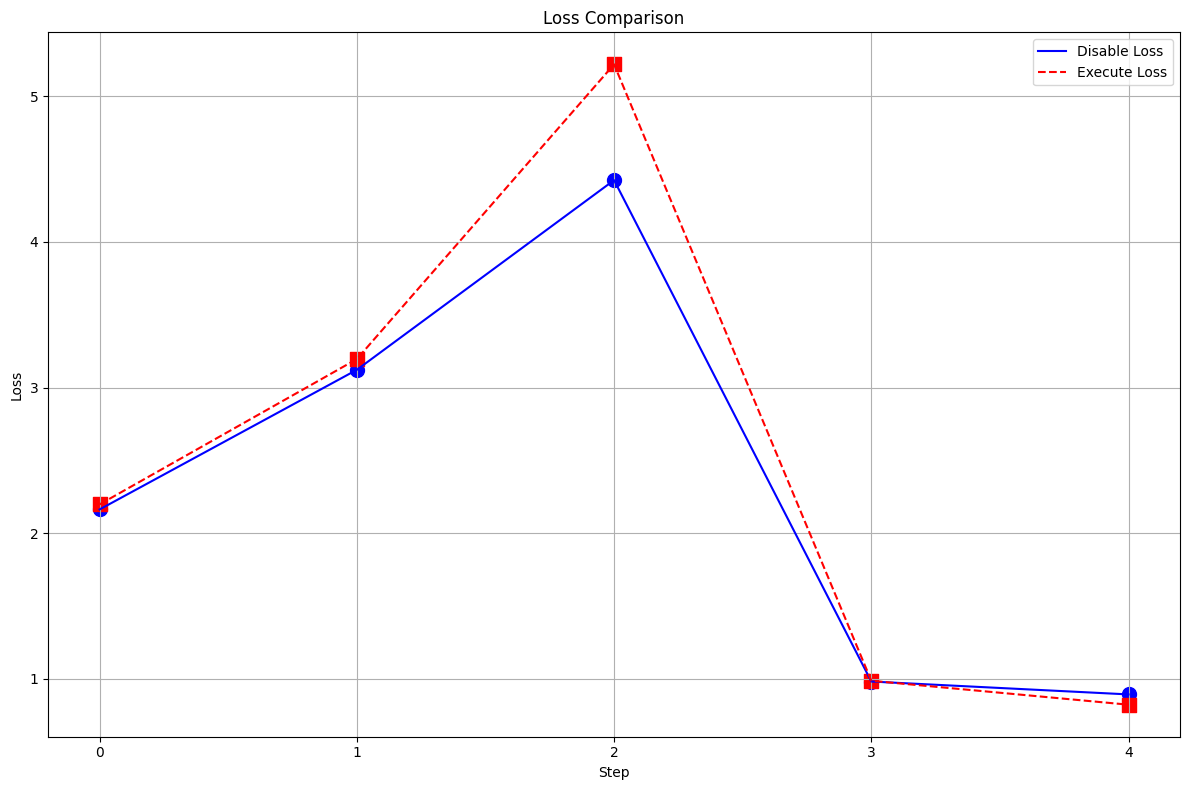}
  \caption{Llama-3.2-1B: training-loss comparison between 8-bit quantized cleartext (blue) and HE execution (red) over the first five steps.}
  \label{fig:loss_fhe_llama1b}
\end{figure}

\paragraph{Training-loss fidelity (Llama-3.2-1B).}
We first contrasted the early loss trajectories of an 8-bit quantized Llama-3.2-1B fine-tuning run executed \emph{in cleartext} against the identical run under HE. Each step invokes multiple encrypted vector-matrix products per transformer layer during both forward and backward passes. Figure~\ref{fig:loss_fhe_llama1b} shows the first five optimization steps: the cleartext (solid blue) and homomorphic (dashed red) curves overlap almost perfectly, with any deviations well within stochastic batching noise. This near-perfect match confirms that ciphertext noise growth, modulus switching, and homomorphic operations do not materially perturb gradient computations.

\paragraph{Bit-level error analysis.}
To complement the loss-based test and precisely quantify any residual numerical errors inherent in the HE computation, we measured the bit error rate in decrypted dot-product results across different bit positions. This analysis simulated the core vector-matrix multiplication by performing homomorphic dot products between encrypted random integer vectors and cleartext random integer weight vectors of varying input dimensions $d_{in}$ (specifically, 768, 2048, and 8192). This dimension represents the input vector size, corresponding to the inner dimension of the weight matrix column involved in the dot product. All these HE computations used the fixed polynomial size $N=2048$ (as defined in Table~\ref{tbl:cryptoparams}). We decrypted the resulting HE ciphertexts and compared them bitwise against the true cleartext dot-product values.

\begin{figure}[H]
  \centering
  \includegraphics[width=0.75\linewidth]{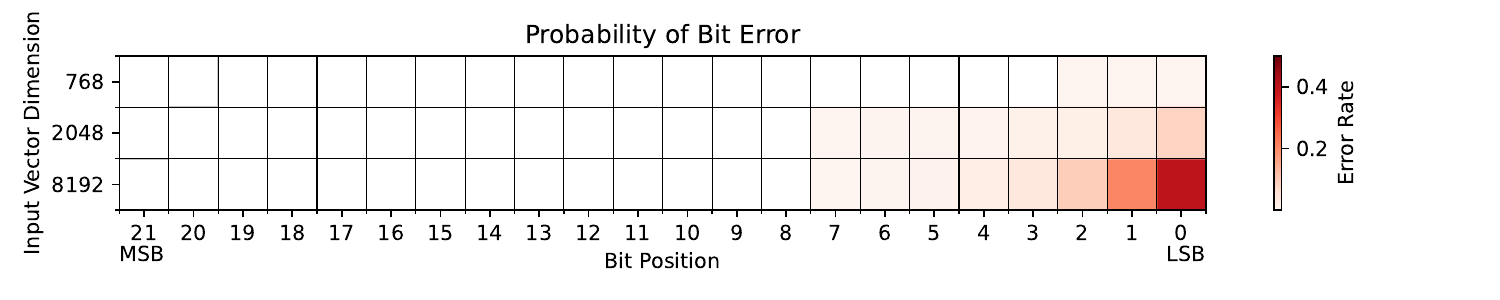}
\caption{Bit error rate versus bit position for homomorphic dot products. The y-axis shows the \textbf{input vector dimension $d_{in}$} (dimensions tested: 768, 2048, 8192). The x-axis shows the bit position (MSB near 21, left; LSB=0, right). All computations used the fixed polynomial size $N=2048$. Higher input dimensions increase LSB error due to noise accumulation, but high-order bits (positions 12 and higher) exhibit error rates under 1\%, preserving $\gamma=12$ MSBs reliably.}
    \label{fig:bit_error}
\end{figure}

Figure~\ref{fig:bit_error} reports the observed error rates. The y-axis represents the input vector dimension $d_{in}$ used in the dot product, while the x-axis shows the bit position, ranging from More Significant Bits (MSBs) on the left (e.g., position 21) down to the Least Significant Bit (LSB) at position 0 on the right. We observe two key points:
\begin{itemize}
    \item \textbf{Noise increases with dimension:} As the input vector dimension $d_{in}$ increases, the error rate in the lower-order bits (LSBs, closer to 0) tends to increase. This is expected because computing the dot product involves accumulating more terms homomorphically, which leads to greater noise accumulation in the HE ciphertext.
    \item \textbf{MSBs remain accurate:} Despite the noise growth affecting the LSBs, the critical high-order bits remain highly accurate across all tested dimensions. Specifically, for bit positions 12 and higher (i.e., the 13th bit up to the MSB, indexed from 0), the observed error rate is negligible (well below 1\%). This confirms that our HE parameters reliably preserve the $\gamma=12$ most significant bits required for computational accuracy, even for the largest input dimension tested ($d_{in}=8192$).
\end{itemize}
This analysis verifies that our cryptographic setup maintains the necessary numerical precision for the core homomorphic operations within the fine-tuning process.

\subsection{HE Performance}
\label{sec:fhe_performance}

We analyze the performance of the HE computations on the server side, focusing first on the core encrypted vector–clear matrix multiplication ($W \cdot [x]_\text{HE}$) and then on the end-to-end time for a full training step, including the implications for client-side computation. All timings were obtained on a single NVIDIA RTX 4060 Laptop GPU using the cryptographic parameters from Table~\ref{tbl:cryptoparams}.

\paragraph{Core HE Operation Performance.}
The server's primary task is computing $W \cdot [x]_\text{HE}$. We benchmarked this operation for typical LLM layer dimensions ($d_{in}, d_{out}$), considering an input batch of encrypted vectors $[x]_\text{HE}$ multiplied by a cleartext weight matrix $W$. Table~\ref{tbl:fhe_matmul_timing} lists the measured latencies for a single token ($B=1, C=1$, where $C$ is context length). The time scales approximately linearly with the total number of input tokens ($B \times C$) and the output dimension ($d_{out}$), and depends on the input dimension $d_{in}$ relative to the polynomial size $N$ used for packing.

\begin{table}[H]
\centering
\caption{Latency of encrypted-vector $\times$ clear-matrix multiplication ($W \cdot [x]_\text{HE}$) for a \emph{single} token ($B=1, C=1$) on an RTX 4060 Laptop GPU. Polynomial size $N=2048$. Reported values are mean $\pm$ std. dev.}
\label{tbl:fhe_matmul_timing}
\begin{tabular}{@{}ccc@{}}
\toprule
Input Dim ($d_{in}$) & Output Dim ($d_{out}$) & Latency (seconds) \\ \midrule
\;768  & \;768 & $0.0809 \pm 0.0011$ \\
3072   & \;768 & $0.1528 \pm 0.0267$ \\
2048   & 2048  & $0.2402 \pm 0.0253$ \\
\;768  & 3072  & $0.3389 \pm 0.0271$ \\
8192   & 2048  & $0.6368 \pm 0.0202$ \\
2048   & 8192  & $1.0539 \pm 0.0335$ \\ \bottomrule
\end{tabular}
\end{table}

\paragraph{End-to-End Training Step Performance.}
To obtain an end-to-end timing for the HE back-end, we ran a \emph{single} training step (forward \emph{and} backward pass) of the \textbf{Llama-3.2-1B} model ($n_{layers}=16, d=2048, m=8192, r=8$) on a code-generation task. This step invokes the $W \cdot [x]_\text{HE}$ primitive for the various weight matrices in each transformer layer. The mini-batch contained $B=1$ sequence, truncated and padded to a context length of $C = 16$ tokens. All linear layers acting on the public base weights $W$ were offloaded to the server and executed under HE.

\paragraph{Raw timing.}
The complete step (forward \emph{and} backward pass) for this batch ($B=1, C=16$) finished in \textbf{57 min 38 s} (\textbf{3458 s}). This corresponds to processing $16$ tokens in $3458$ seconds, yielding an average throughput of approximately $0.0046$ tokens per second, or a latency of:
\[
  \tau_{\text{tok}} \approx \frac{3458\,\text{s}}{16\,\text{tokens}} \approx \textbf{216}\;\text{s per token}.
\]

\paragraph{Scaling and Parallelization Potential.}
The total wall-time $T$ for an HE training step is expected to scale linearly with the total number of encrypted tokens processed: $T \approx \tau_{\text{tok}}\,(B\,C)$. Consequently, doubling the batch size or the sequence length would roughly double the latency on a single server. For inference (forward pass only), the cost would drop by approximately a factor of 2. Crucially, because the client-side workload is minimal (see below) and the server computations for different tokens or batches can often be parallelized, this approach exhibits \textbf{excellent scaling potential}. The overall throughput can be significantly increased by distributing the workload across $S$ identical HE servers, potentially reducing the effective per-token latency towards $\tau_{\text{tok}}/S$.

\paragraph{Client Compute Rate.}
Using the derived formula (Eq.~\eqref{eq:client_comp}, using the approx. $2 \times$ factor for backward pass) and the Llama-3.2-1B parameters ($n_{layers}=16, d=2048, m=8192, r=8, C=16$), we estimate the client's computational work during this benchmark step:
\begin{align*}
\textbf{Client}_{FLOPs} &\approx 2 \times 16 \times (2 \times 2048 \times 16^2 + 22 \times 2048 \times 8 + 6 \times 8192 \times 8) \\
&\approx 48 \times (1,048,576 + 360,448 + 393,216) \\
&\approx 48 \times 1,802,240 \approx 86.5 \times 10^6 \text{ FLOPs}.
\end{align*}
Given the total step time of $3458$\,s, the average client compute rate required is:
\[
  \frac{86.5 \times 10^6 \text{ FLOPs}}{3458 \text{ s}} \approx 25,000 \text{ FLOP/s} \approx \textbf{0.025 MFLOP/s}.
\]
This extremely low rate confirms that the client's computational burden is negligible. The overall process is heavily bottlenecked by the HE computations on the server, reinforcing the viability of using multiple parallel servers managed by a lightweight client.

\paragraph{Resource footprint.}
During the benchmark, the server held at most $\approx 2.8$ GB of RLWE ciphertexts, comfortably within the 8 GB VRAM of the RTX 4060 Laptop GPU. Using the expansion factors from Sec.~4.4, we estimate the data transfer \textit{per HE-accelerated linear layer invocation} for the batch ($B=1, C=16$). Assuming $d_{in}=d_{out}=2048$, $N=2048$, so $L=L'=1$ block per token:
\begin{itemize}
    \item Client to Server (Input Activation Ciphertexts): $16 \text{ tokens} \times 1 \text{ block/token} \times 9992 \text{ bytes/block} \approx 160$ kB.
    \item Server to Client (Output Activation Ciphertexts): $16 \text{ tokens} \times 1 \text{ block/token} \times 6656 \text{ bytes/block} \approx 107$ kB.
\end{itemize}
Total transfer per layer invocation $\approx 267$ kB. A full training step involves multiple such invocations (for different layers $W_Q, W_K, \dots$ and for the backward pass). Even accounting for $n_{layers} \times (\text{\# HE layers}) \times 2$ transfers, the total bandwidth per step remains modest, well within the capacity of standard internet connections. (Note: Bandwidth scales linearly with $B \times C$ and the number of HE layers.)

\subsection{Qualitative Results}
\label{sec:qualitative}
We demonstrate the effect of fine-tuning by comparing model outputs before and after training (using the converged quantized cleartext models as a proxy for HE results).

\textbf{Llama-3.2-1B on Code Generation:} We prompted the Llama-3.2-1B model before and after fine-tuning on Python snippets related to the Concrete ML library. Listing~\ref{lst:llama_base} shows the base model providing generic Python parameters typical for standard machine learning libraries. In contrast, Listing~\ref{lst:llama_tuned} demonstrates the fine-tuned model correctly suggesting the \texttt{n\_bits} argument, which is specific to the Concrete ML library and reflects the context provided during fine-tuning. This illustrates successful adaptation to a specialized technical domain.

\begin{lstlisting}[caption={Llama-3.2-1B Original model prediction (Code Gen)},label={lst:llama_base},captionpos=b]
Prompt: from concrete.ml.sklearn import LogisticRegression\n\nmodel = LogisticRegression(
Completion: parameters={'C': [0.1, 1, 10], 'penalty': ['l1', 'l2']}) # Example standard parameters
\end{lstlisting}

\begin{lstlisting}[caption={Llama-3.2-1B Fine-tuned model prediction (Code Gen, 8-bit quantized)},label={lst:llama_tuned},captionpos=b]
Prompt: from concrete.ml.sklearn import LogisticRegression\n\nmodel = LogisticRegression(
Completion: n_bits=8) # Example Concrete ML specific parameter
\end{lstlisting}

\textbf{Llama-3.2-1B on Mathematical Reasoning (Orca-Math):} We also tracked the model's ability to solve simple mathematical word problems during fine-tuning on the Orca-Math dataset. The prompt "When you multiply a number by 7, it becomes 98. What is that number?" was taken from the test-set and used for evaluation at various checkpoints.

Early in training (e.g., Step 100, Listing~\ref{lst:llama_math_early}), the model exhibited confusion, often misinterpreting the question or performing incorrect operations:
\begin{lstlisting}[caption={Llama-3.2-1B at Step 100 (Orca-Math, 8-bit quantized)},label={lst:llama_math_early},captionpos=b]
Prompt: When you multiply a number by 7, it becomes 98. What is that number?
Response: If you multiply a number by 7, it becomes 98. So, the number you're asking about is 98.
\end{lstlisting}

After sufficient fine-tuning (e.g., Step 1000, Listing~\ref{lst:llama_math_late}), the model consistently demonstrated the correct reasoning process—identifying the need for division and performing the calculation accurately:
\begin{lstlisting}[caption={Llama-3.2-1B at Step 1000 (Orca-Math, 8-bit quantized)},label={lst:llama_math_late},captionpos=b]
Prompt: When you multiply a number by 7, it becomes 98. What is that number?
Response: To find the number, you need to divide 98 by 7.

98 / 7 = 14

So the number is 14
\end{lstlisting}

These examples illustrate that the private fine-tuning process, even when employing HE-compatible quantization, effectively adapts the models to the specific nuances, terminology, and problem-solving skills required by the target domain specified by the private data.

\section{Conclusion}

We presented and validated an interactive protocol using Homomorphic Encryption (HE) to enable privacy-preserving LoRA fine-tuning of open-source LLMs. By strategically outsourcing computations involving the public base model weights to an HE-enabled server while keeping private data and LoRA adapters on the client, our approach addresses the confidentiality challenge inherent in adapting LLMs to sensitive domains.

We demonstrated the feasibility of this method by fine-tuning the Llama-3.2-1B model. Our experiments confirmed that carefully chosen HE-compatible quantization can achieve convergence nearly identical to floating-point training, and the HE execution faithfully replicates the quantized cleartext dynamics. Performance benchmarks on a single GPU, while indicating significant computational cost ($\approx$ 216 s/token for a Llama-3.2-1B training step), highlighted the viability of our optimized HE implementation and demonstrated the extremely low client-side compute requirement ($\approx$ 0.025 MFLOP/s). This low client burden, combined with modest bandwidth needs, makes multi-server parallelization a practical and promising approach for scaling the HE computation to achieve acceptable training times.

This work provides a concrete pathway for securely leveraging private datasets to specialize LLMs in fields like healthcare or finance. While performance optimization and scaling remain crucial, our results show the potential of HE to unlock privacy-sensitive AI applications. Future work includes further HE optimizations, exploring efficient multi-server orchestration, and extending the approach to embedding layers.

% \paragraph{Code Availability.} The code implementing our protocol and experiments is available as part of the Concrete ML library: \url{https://github.com/zama-ai/concrete-ml}. Example notebooks reproducing the experiments will be made available upon publication at [Link/Repository].

% \clearpage
\bibliographystyle{splncs03}
\bibliography{main}

\begin{thebibliography}{10}
\providecommand{\url}[1]{\texttt{#1}}
\providecommand{\urlprefix}{URL }

\bibitem{GEMM}
How to optimize a cuda matmul kernel for cublas-like performance: a worklog. \url{https://siboehm.com/articles/22/CUDA-MMM}, accessed: 2024-10-30

\bibitem{llama32024}
AI, M.: The llama 3 herd of models. arXiv preprint arXiv:2407.21783  (2024), \url{https://arxiv.org/abs/2407.21783}

\bibitem{bergamaschi2019homomorphic}
Bergamaschi, F., Halevi, S., Halevi, T.T., Hunt, H.: Homomorphic training of 30,000 logistic regression models. In: Applied Cryptography and Network Security: 17th International Conference, ACNS 2019, Bogota, Colombia, June 5--7, 2019, Proceedings 17. pp. 592--611. Springer (2019)

\bibitem{Bonte2018}
Bonte, C., Vercauteren, F.: Privacy-preserving logistic regression training. BMC Medical Genomics  11(4), ~86 (Oct 2018), \url{https://doi.org/10.1186/s12920-018-0398-y}

\bibitem{TFHE}
Chillotti, I., Gama, N., Georgieva, M., Izabach{\`{e}}ne, M.: {TFHE:} fast fully homomorphic encryption over the torus. Journal of Cryptology  33(1),  34--91 (2020)

\bibitem{dettmers2024qlora}
Dettmers, T., Pagnoni, A., Holtzman, A., Zettlemoyer, L.: Qlora: Efficient finetuning of quantized llms. Advances in Neural Information Processing Systems  36 (2024)

\bibitem{gao2024dlora}
Gao, C., Zhang, S.Q.: Dlora: Distributed parameter-efficient fine-tuning solution for large language model. arXiv preprint arXiv:2404.05182  (2024)

\bibitem{mpc_dot_prod}
Goethals, B., Laur, S., Lipmaa, H., Mielik\"{a}inen, T.: On private scalar product computation for privacy-preserving data mining. p. 104–120. ICISC'04, Springer-Verlag, Berlin, Heidelberg (2004), \url{https://doi.org/10.1007/11496618_9}

\bibitem{han2019logistic}
Han, K., Hong, S., Cheon, J.H., Park, D.: Logistic regression on homomorphic encrypted data at scale. In: Proceedings of the AAAI conference on artificial intelligence. vol.~33, pp. 9466--9471 (2019)

\bibitem{hao2022iron}
Hao, M., Li, H., Chen, H., Xing, P., Xu, G., Zhang, T.: Iron: Private inference on transformers. Advances in neural information processing systems  35,  15718--15731 (2022)

\bibitem{hu2022lora}
Hu, E.J., Shen, Y., Wallis, P., Allen-Zhu, Z., Li, Y., Wang, S., Wang, L., Chen, W.: Lo{RA}: Low-rank adaptation of large language models. In: International Conference on Learning Representations (2022), \url{https://openreview.net/forum?id=nZeVKeeFYf9}

\bibitem{Cheetah}
Huang, Z., jie Lu, W., Hong, C., Ding, J.: Cheetah: Lean and fast secure {Two-Party} deep neural network inference. In: 31st USENIX Security Symposium (USENIX Security 22). pp. 809--826. USENIX Association, Boston, MA (Aug 2022), \url{https://www.usenix.org/conference/usenixsecurity22/presentation/huang-zhicong}

\bibitem{jacob2018quantization}
Jacob, B., Kligys, S., Chen, B., Zhu, M., Tang, M., Howard, A., Adam, H., Kalenichenko, D.: Quantization and training of neural networks for efficient integer-arithmetic-only inference. In: Proceedings of the IEEE conference on computer vision and pattern recognition. pp. 2704--2713 (2018)

\bibitem{10.1007/978-3-031-30872-7_5}
Joye, M.: On-line/off-line dcr-based homomorphic encryption and applications. In: Rosulek, M. (ed.) Topics in Cryptology -- CT-RSA 2023. pp. 115--131. Springer International Publishing, Cham (2023)

\bibitem{Gazelle}
Juvekar, C., Vaikuntanathan, V., Chandrakasan, A.: {GAZELLE}: A low latency framework for secure neural network inference. In: 27th USENIX Security Symposium (USENIX Security 18). pp. 1651--1669. USENIX Association, Baltimore, MD (Aug 2018), \url{https://www.usenix.org/conference/usenixsecurity18/presentation/juvekar}

\bibitem{Kim2018}
Kim, A., Song, Y., Kim, M., Lee, K., Cheon, J.H.: Logistic regression model training based on the approximate homomorphic encryption. BMC Medical Genomics  11(4), ~83 (Oct 2018), \url{https://doi.org/10.1186/s12920-018-0401-7}

\bibitem{lou2020glyph}
Lou, Q., Feng, B., Charles~Fox, G., Jiang, L.: Glyph: Fast and accurately training deep neural networks on encrypted data. Advances in neural information processing systems  33,  9193--9202 (2020)

\bibitem{lyubashevsky2010rlwe}
Lyubashevsky, V., Peikert, C., Regev, O.: On ideal lattices and learning with errors over rings. In: Advances in Cryptology -- EUROCRYPT 2010. Lecture Notes in Computer Science, vol. 6110, pp. 1--23. Springer (2010)

\bibitem{10.1145/3643651.3659891}
Montero, L., Frery, J., Kherfallah, C., Bredehoft, R., Stoian, A.: Machine learning training on encrypted data with tfhe. In: Proceedings of the 10th ACM International Workshop on Security and Privacy Analytics. p. 71–76. IWSPA '24, Association for Computing Machinery, New York, NY, USA (2024), \url{https://doi.org/10.1145/3643651.3659891}

\bibitem{nandakumar2019towards}
Nandakumar, K., Ratha, N., Pankanti, S., Halevi, S.: Towards deep neural network training on encrypted data. In: Proceedings of the IEEE/CVF conference on computer vision and pattern recognition workshops. pp. 0--0 (2019)

\bibitem{Paillier}
Paillier, P.: Public-key cryptosystems based on composite degree residuosity classes. In: Stern, J. (ed.) Advances in Cryptology --- EUROCRYPT '99. pp. 223--238. Springer Berlin Heidelberg, Berlin, Heidelberg (1999)

\bibitem{regev2005lwe}
Regev, O.: On lattices, learning with errors, random linear codes, and cryptography. In: Proceedings of the 37th Annual ACM Symposium on Theory of Computing (STOC 2005). pp. 84--93. ACM (2005)

\bibitem{wang2023privatelora}
Wang, Y., Lin, Y., Zeng, X., Zhang, G.: Privatelora for efficient privacy preserving llm. arXiv preprint arXiv:2311.14030  (2023)

\bibitem{xiao2023smoothquant}
Xiao, G., Lin, J., Seznec, M., Wu, H., Demouth, J., Han, S.: Smoothquant: Accurate and efficient post-training quantization for large language models. In: International Conference on Machine Learning. pp. 38087--38099. PMLR (2023)

\bibitem{yang2024packvflefficientpackingvertical}
Yang, L., Cai, S., Chai, D., Zhang, J., Tian, H., Jin, Y., Guo, K., Chen, K., Yang, Q.: Packvfl: Efficient he packing for vertical federated learning (2024), \url{https://arxiv.org/abs/2405.00482}

\bibitem{BatchCrypt}
Zhang, C., Li, S., Xia, J., Wang, W., Yan, F., Liu, Y.: {BatchCrypt}: Efficient homomorphic encryption for {Cross-Silo} federated learning. In: 2020 USENIX Annual Technical Conference (USENIX ATC 20). pp. 493--506. USENIX Association (Jul 2020), \url{https://www.usenix.org/conference/atc20/presentation/zhang-chengliang}

\end{thebibliography}

\end{document}